# Effect of Mg and C contents in $MgCNi_3$, and structure and superconductivity of $MgCNi_{3-x}Co_x$

Z.A. Ren[*], G.C. Che, S.L. Jia, H. Chen, Y.M. Ni, Z.X. Zhao

National Laboratory for superconductivity, Institute of Physics

Chinese Academy of Sciences, P.O. Box 603, Beijing 100080, P.R. China

Abstract:

The effect of Mg and C contents on $T_C$ in $MgCNi_3$, and structure and superconductivity of $MgCNi_{3-x}Co_x$ were studied. It is found that the excess of Mg and C in initial material mixture is favorable to improve $T_C$ and obtain single-phase samples. For preparing $MgCNi_3$ superconductor, the optimum composition of starting materials is $MgC_{1.45}Ni_3$ with 20wt.% excess of Mg of the stoichiometric composition. In $MgCNi_{3-x}Co_x$ system, a continuous solid solution is formed, lattice parameter decreases slightly and $T_C$ decreases obviously with increasing x. A suppression of superconductivity is observed due to the substitution of Co (Mn) for Ni. The suppression effect is smaller for the substitution of Co than that of Mn.

---

[*] Corresponding author: renzhian@ssc.iphy.ac.cn

I. Introduction

The recent discovery of superconductivity at 40K in $MgB_2$ [1] shows that intermetallic compounds are worth of reconsideration to search for new superconducting materials. After the discovery of $MgB_2$, another intermetallic compound superconductor, $MgCNi_3$ with $T_C$ = 8K, was found [2]. This material is interest due to its simple perovskite structure and nickel-rich composition that usually results in magnetism. In this paper, the effect of Mg and C contents on $T_C$ in $MgCNi_3$, and the structure and superconductivity of $MgCNi_{3-x}Co_x$ alloy are studied.

II. Experiment

Alloy samples were prepared by powder metallurgy method. The fine powders of Mg, C, Ni and Co with purity better than 99.5% were used as the starting materials. The mixture with appropriate composition was pressed into pellets. The pellets were wrapped with Ta foil and enclosed in an evacuated quartz tube, then placed in a box furnace and heated to 950°C at a rate of 150°C/h and kept at this temperature for 5h, followed by furnace-cooling to room temperature. The surface layer of sintered samples was polished off. X-ray diffraction (XRD) analysis was performed by an M18X-AHF type diffractometer with Cu-$K_\alpha$ radiation. PowderX and LAZY programs were used for lattice parameter calculations. Resistance curves and magnetization were measured by the standard four-probe technique and a DC-SQUID magnetometer, respectively.

III. Results and Discussion

1. Effect of Mg and C contents in $MgCNi_3$

In order to investigate the effect of Mg and C contents on $T_C$ in $MgCNi_3$ and obtain single-phase sample, $MgC_xNi_3$ samples with different C content (x = 1.25, 1.35, 1.45, and 1.55) and $Mg_{1+\delta}C_{1.45}Ni_3$ samples with different $\delta$ ($\delta$ = 0, 10, 20, 30wt.%) were prepared, where δ is the Mg excess of the stoichiometric composition employed in the initial mixture.

Our results indicate that the $T_C$ of $MgC_xNi_3$ is sensitive to the C content. The highest $T_C$ and the best superconducting transition can be obtained when C content x = 1.45 as shown in figure 1, which is in agreement with that of Ref. [2]. The FC and ZFC magnetization curves of $MgC_{1.45}Ni_3$ with nominal composition are shown in figure 2.

We find that Mg excess is favorable to improve $T_C$ and obtain single-phase sample due to the volatility of Mg. Figure 3 shows the temperature dependence of resistivity of samples with different excess of Mg and indicates that the sample with Mg excess of 20wt.% has the best superconducting transition and the highest $T_C$(zero) temperature. Therefore, Mg excess of 20wt.% was employed for the preparation of $MgC_{1.45}Ni_{3-x}Co_x$ alloys.

By the way, we point out that in $MgC_{1.45-x}B_xNi_3$ alloys with x = 0.05, 0.10, 0.15, and 0.20, single-phase samples can be obtained with x = 0.0 – 0.1, but the $T_C$ of these samples decreases. For the samples with x = 0.15 – 0.20, single-phase samples cannot be obtained and superconductivity of these samples cannot be observed by resistance measurements.

2. Structure of $MgC_{1.45}Ni_{3-x}Co_x$

The structure of nominal composition $MgC_{1.45}Ni_3$ has been determined by powder neutron diffraction at ambient temperature [2, 3, 5]. The compound has a simple cubic perovskite structure of $CaTiO_3$ type with space group Pm-3m. The positions for the atoms are Mg: 1a (0,0,0), C: 1b (0.5,0.5,0.5) and Ni: 3c (0,0.5,0.5). The C site occupancy was found to be 0.960(8). The actual formula for the superconducting phase was determined to be $MgC_{0.96}Ni_3$. To date, it is not clear that why carbon must be excess in initial material mixtures for obtaining good quality samples. A small amount of unreacted carbon can be detected by STM and EDS observations, but it is difficult to detect carbon by XRD.

Typical XRD patterns of nominal composition of $MgC_{1.45}Ni_{3-x}Co_x$ are shown in figure 4. It shows clearly that the samples are single-phase while Co content x increases from 0 to 3. The change of lattice parameters in this system is shown in figure 5. The refined lattice parameter is a = 3.806(4) Å for $MgC_{1.45}Ni_3$ and a = 3.802(6) Å for $MgC_{1.45}Co_3$, respectively. Figure 5 shows that lattice parameters of $MgC_{1.45}Ni_{3-x}Co_x$ decrease slightly. For a direct understanding, structure of $MgCNi_3$ is shown in figure 6.

3. Superconductivity in $MgC_{1.45}Ni_{3-x}Co_x$

Figure 7 presents the R-T curves of $MgC_{1.45}Ni_{3-x}Co_x$. It indicates that the superconducting transition temperature $T_C$ decreases gradually with increasing x. The resistance can not reach zero for the samples with Co content x > 0.50 at low temperature. A small superconducting transition onset can be still observed for the sample with x = 1.5, this indicates that the superconducting volume fraction decreases due to the substitution of Co for Ni, which is confirmed by magnetization measurements. Our results are different from that of reference [4]. To confirm that the observed superconductivity for the samples with x > 0.5 is not from the inhomogeneity of the small amount Co distribution in $MgC_{1.45}Ni_3$, a contrast experiment has been done. $MgC_{1.45}Ni_{3-x}Mn_x$ samples with x = 0.125 and 0.5 prepared by the same condition as that of $MgC_{1.45}Ni_{3-x}Co_x$ are also single-phase, but their superconductivity can not be observed by resistance measurements. This indicates that the substitution of Co for Ni leads to a suppression of superconductivity and the suppression effect is smaller than that of Mn for Ni.

In summary, the effect of Mg and C contents on $T_C$ in $MgCNi_3$, and the structure and superconductivity of $MgCNi_{3-x}Co_x$ system were studied. It is found that the excess of Mg and C in initial material mixture is favorable to improve $T_C$ and obtain single-phase samples. For preparing $MgCNi_3$ superconductor, the optimum composition of starting materials is $MgC_{1.45}Ni_3$ with the Mg excess 20wt.% of stoichiometric composition. In $MgCNi_{3-x}Co_x$ system, a continuous solid solution is formed, superconducting transition temperature decreases with increasing x. A suppression of superconductivity is observed due to the substitution of Co (Mn) for Ni. The suppression effect of the substitution of Co is smaller than that of Mn.

Acknowledgement:

This work is supported by the Ministry of Science and Technology of China (NKBRSF-G 19990646).

**Figure captions:**

Fig.1. Resistance vs. Temperature curves of $MgC_xNi_3$

Fig.2. Zero-field-cooled (ZFC) and field-cooled (FC) DC magnetization curves of nominal composition $MgC_{1.45}Ni_3$

ZFC curves was obtained by Zero-field cooling the sample to 4K, then applying an external field of 10Oe, and subsequently warming up to 300K, FC curves was obtained by applying an external field of 10Oe, then cooling the sample to 4K, and subsequently warming up to 300K

Fig.3. R-T curves of $Mg_{1+\delta}C_{1.45}Ni_3$ with different excess of Mg

Fig.4. Typical XRD patterns of nominal composition $MgC_{1.45}Ni_{3-x}Co_x$

Fig.5. Lattice parameter vs. x in nominal composition $MgC_{1.45}Ni_{3-x}Co_x$

Fig.6. The perovskite crystal structure of $MgCNi_3$ and its projections on (100) and (111) planes

Fig.7. R-T curves of nominal composition $MgC_{1.45}Ni_{3-x}Co_x$

Figure. 1

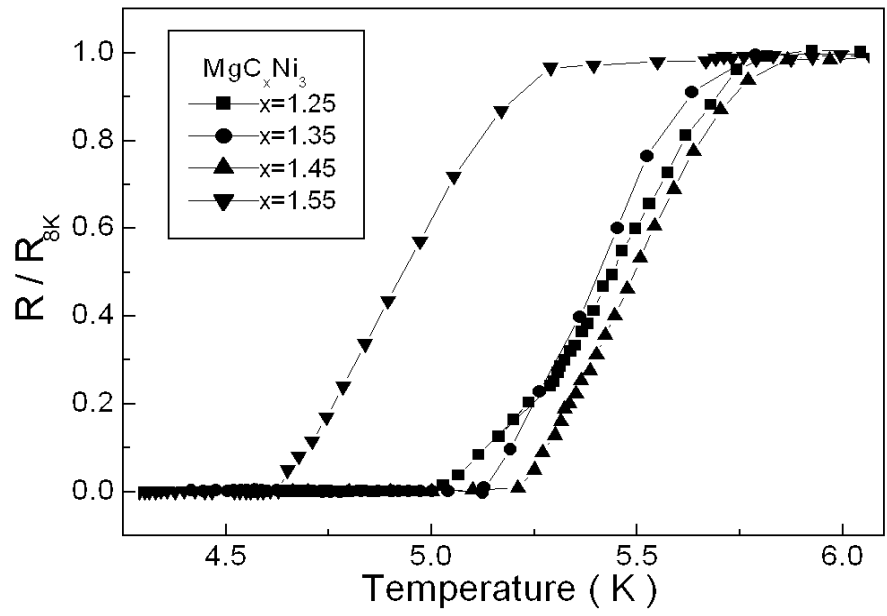

Figure. 2

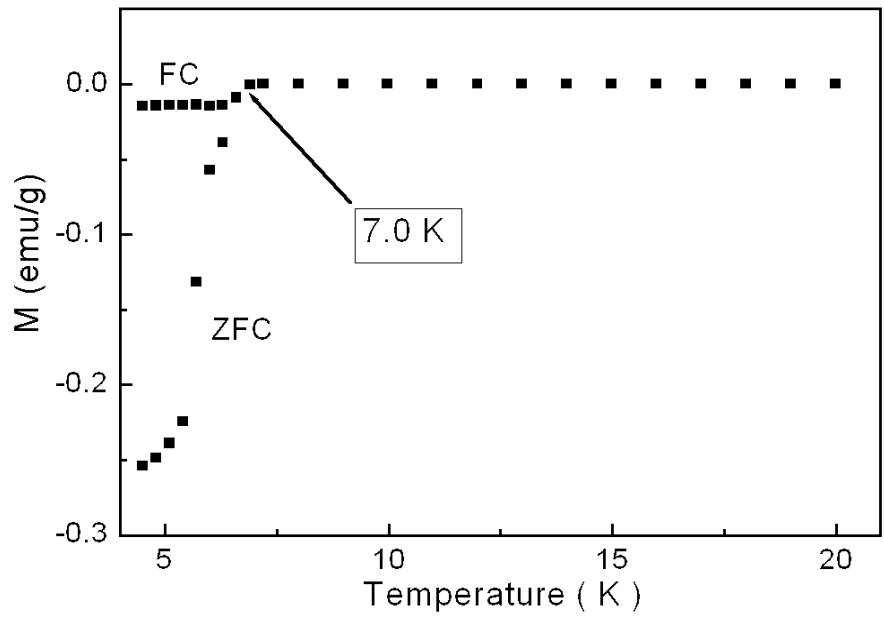

Figure. 3

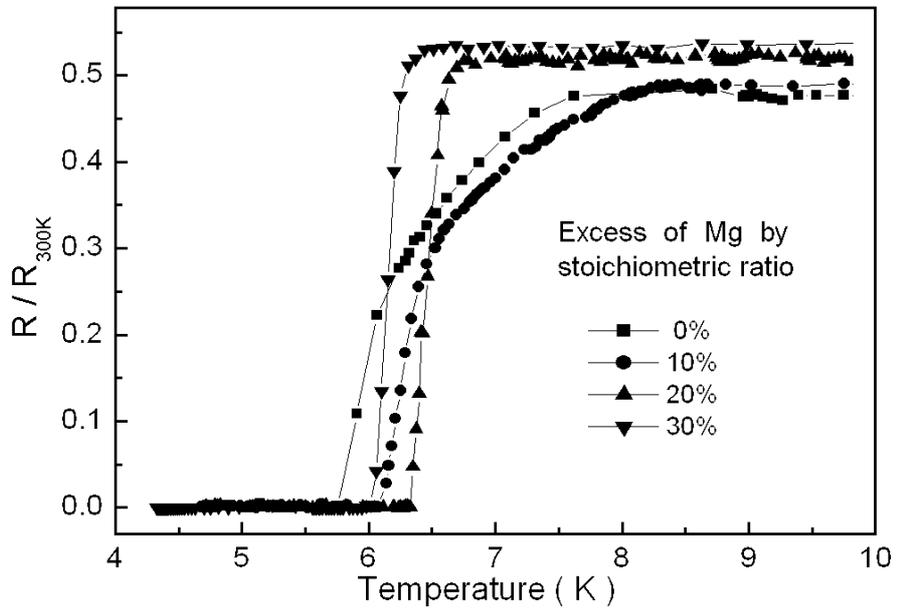

Figure. 4

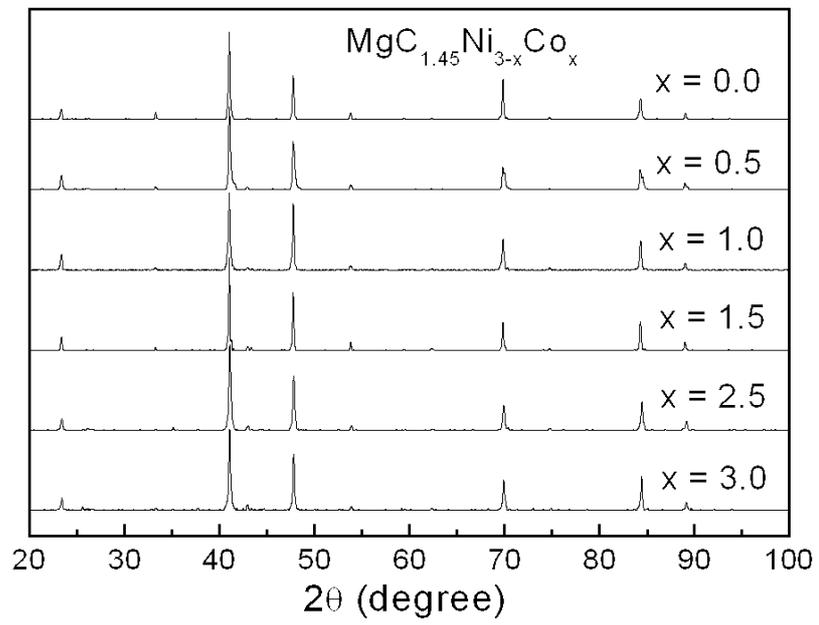

Figure. 5

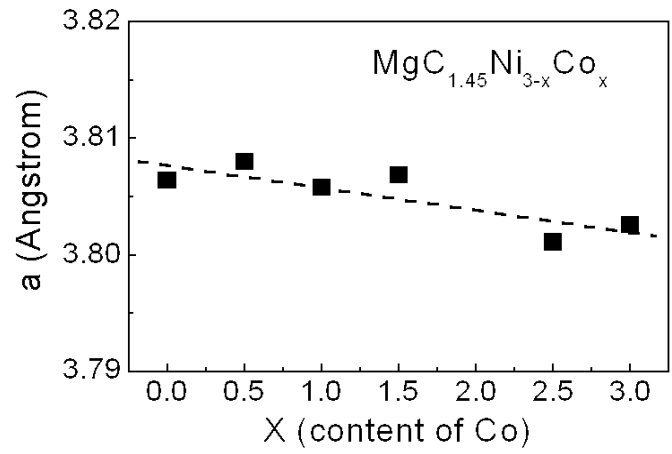

Figure. 6

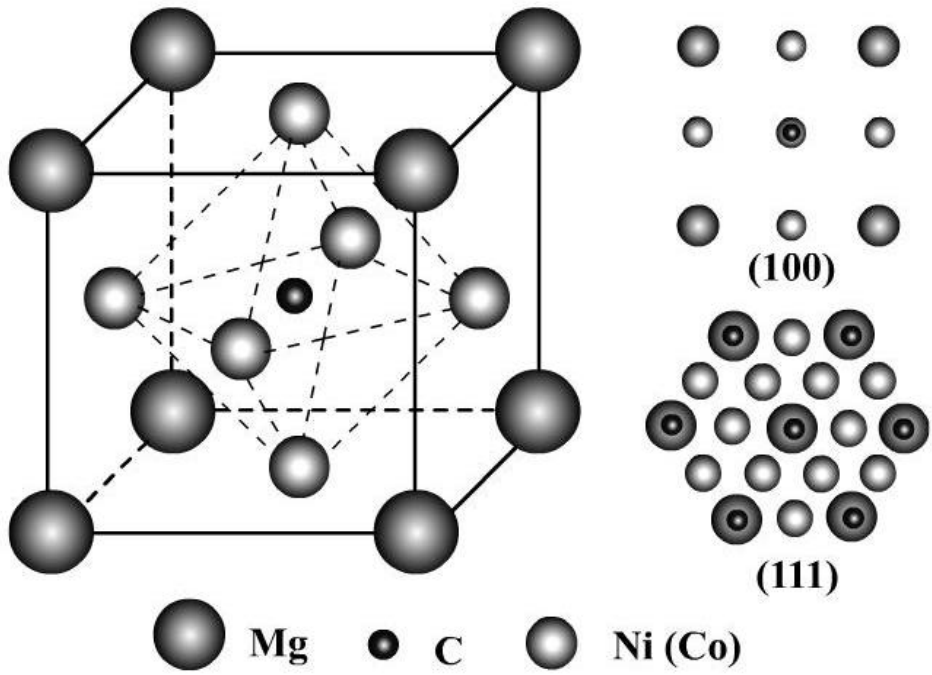

Figure. 7

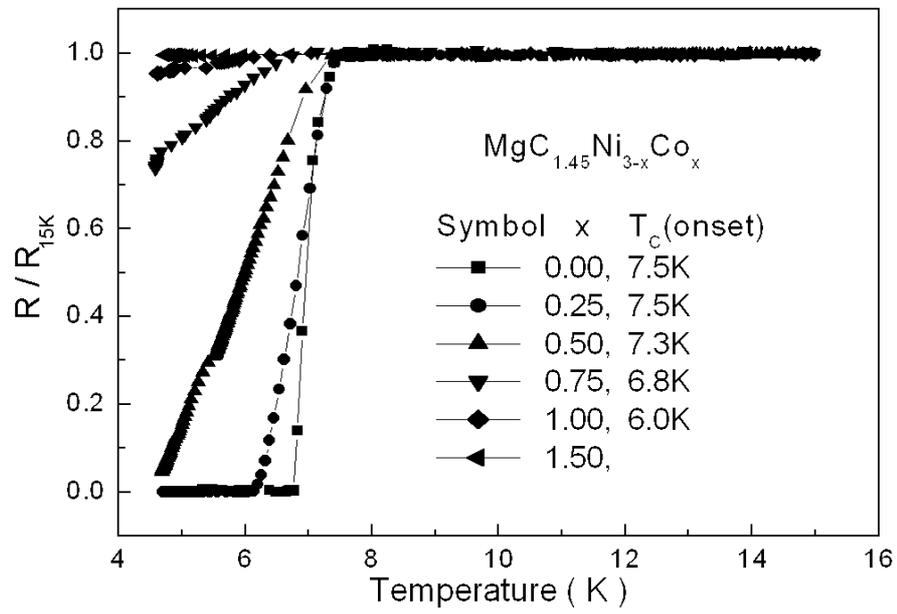